\documentclass[12pt]{article}

\usepackage{jheppub}

\usepackage{amssymb,amsmath}
\usepackage{mathrsfs}
\usepackage{natbib} 

\def\be{\begin{equation}}
\def\ee{\end{equation}}
\def\beqa{\begin{eqnarray}}
\def\eeqa{\end{eqnarray}}
\def\P{\mathscr{P}}

\def\Tr{\mbox{Tr}}
\def\rra{\rangle\rangle}

\def\ln{\text{ln}}
\def\D{{\cal D}}
\def\hD{\hat{\cal D}}

\def\S{\hat{S}}
\def\bd{\breve{d}}
\def\m{\text{min}}

\begin{document}

\begin{center}
\Large{\bf On the topological  and crosscap entropies in non-oriented RCFTs}
\vspace{0.5cm}

\large  Hugo Garc\'{\i}a-Compe\'an$^a$\footnote{e-mail address: {\tt
compean@fis.cinvestav.mx}}, Norma Quiroz$^{b}$\footnote{e-mail
address: {\tt norma.quiroz@cusur.udg.mx}}

{\small \em $^a$Departamento de F\'{\i}sica, Centro de
Investigaci\'on y de Estudios Avanzados del IPN}\\
{\small\em P.O. Box 14-740, CP. 07000, M\'exico City, M\'exico}\\
\vspace*{0.5cm}
{\small \em $^b$Divisi\'on de Ciencias Exactas, Naturales y Tecnol\'ogicas\\
Centro Universitario del Sur, Universidad de Guadalajara}\\
{\small\em Enrique Arreola Silva No. 883, Centro, 49000 Cd Guzman,
Jal, M\'exico}

\vspace*{1.5cm}
\today
\end{center}

\begin{abstract}

We establish a relation between the boundary and the topological
entropies for the conformal minimal models in some of the simplest
models of the unitary A-A series. We show that in these models the
boundary entropy is a difference of topological entropies.
Furthermore, we define the crosscap entropy as the analog to the
boundary entropy  in non-oriented theories. The crosscap entropy is
defined as the logarithm of  the degeneracy of the ground state due
to the presence of  crosscaps and it can be expressed in terms of
the crosscap coefficients. This crosscap entropy has not an explicit
relation to the topological entropy as the boundary entropy.
However, we propose a new quantity, $\hat S = \ln P_{0i}$   defined
in terms of the modular transformation $P$ between the open and
closed channel of the M\"obius partition function.  With this
quantity, the crosscap entropy can be regarded as a difference of
entropies, very similar to the boundary entropy. We also compute the
left-right entanglement entropy (LREE) for crosscap states and we
express it in terms of $\hat S$. An explicit example of the LREE of
the crosscap state in Wess-Zumino-Witten is carried out. \vskip
1truecm

\end{abstract}

\bigskip
\newpage

\section{Introduction}
\label{sec:intro}

Entanglement is a quantum feature that has become central in studies
of several  phenomena in string theory, quantum chaos and condensed
matter physics. It has recently been observed experimentally that
there are new phase transitions of matter that cannot be explained
with the theoretical framework of the symmetries of the Hamiltonian.
Several investigations point out that these new phase transitions
have their explanation in terms of the topological order.
Topological order can be manifested in several ways, in particular
as the entanglement of the ground state of the Hamiltonian in the
limit of large system size. In a many-body system, this entanglement
is quantified by the topological (entanglement) entropy $S^{top}=
\ln (d_i/{\cal D})$, where $d_i$ is the quantum dimension of a
particle of charge $i$  and ${\cal D}$ is the total quantum
dimension of the medium \cite{Kitaev:2005dm,Levin:2006zz}. In
unitary 1+1 Conformal Field Theory (CFT), these quantum numbers are
defined in terms of the modular-$S$ matrix, and $i$ labels the kind
of primary operator.

When boundary conditions are implemented into the CFT, there is an
entropy associated with the ground state degeneracy $g$ due to the
presence of the boundaries. Specifically, for a 1+1 Boundary CFT
(BCFT) defined on a cylinder of length $L$ and temperature
$T=1/\beta$, with a boundary  $a$ at one of the ends of the
cylinder, the free energy has the contribution $\ln \, Z = \ln\, g_a$
at the large system limit  $L\gg \beta$. Therefore, the boundary
entropy associated to the universal non-integer ground states
degeneracy at the boundary $a$ is  $S^B_a = \ln \, g_a$
\cite{Affleck:1991tk}. Because of the open- closed duality,  $g$ is
expressed in two equivalent forms, in the closed or the open
channel. In the closed channel, $g$ is determined by the
coefficients of the boundary states. Along this work we consider
only Cardy's boundary states, where the boundary index $a$ agrees
with the index of the representations $i$. In this case, the
boundary coefficients are expressed in terms of the $S$-matrix and
hence in terms of the quantum dimensions.

Recently, another entropy has been defined in 1+1 BCFT, this is the
left-right entanglement entropy (LREE). It measures the amount of
quantum correlation between the left (holomorphic) and right
(anti-holomorphic) modes of the CFT.  We recall that in this entropy
the division of the system is not geometric but it is done as a
factorization of the Hilbert space like a tensor product of
subspaces.  Another entropy based on  the decomposition of the
Hilbert space is the momentum-space entanglement given in
\cite{Balasubramanian:2011wt}. The same idea of non-geometrical
division  was also applied to study entanglement entropy for two
conformal field theories separated by conformal interfaces
\cite{Brehm:2015lja,Gutperle:2017enx}. The LREE  was first studied
for bosonic boundary states in  \cite{PandoZayas:2014wsa}  and later
for D-branes in \cite{Zayas:2016drv}. Furthermore, this entropy was
explored in  \cite{Das:2015oha} for boundary states in
$1+1$-Rational Conformal Field Theories. Also LREE for  $1+1$
Wess-Zumino-Witten (WZW) models and the consequences of the
level-rank duality on this entropy  was determined in
\cite{Schnitzer:2015gpa}.  One of the interesting issues of this
entropy is that  the LREE for an Ishibashi state of a diagonal CFT
over a circle of circumference $\ell$ was found to be
$\displaystyle{S = \pi c{\ell}/24\varepsilon  +
\mbox{log}({d}_j/{\cal D})},$ where $\varepsilon$ is a UV cut-off.
Therefore, the LREE of a 2-$d$ CFT can be useful to study the
entanglement entropy of   $2+1$  Topological Field Theories through
a bulk-boundary correspondence \cite{Das:2015oha,Qi2012}. Further
generalizations involving Hilbert spaces of the $2+1$ theory
constructed from states in an arbitrary superposition of Ishibashi
states was proposed in  \cite{Wen:2016snr}.

Entanglement is a quantum feature that has become central. Until
now, the topological and the boundary entropy, as well as the  LREE,
have been studied for CFT's in the cylinder which is an oriented
surface. By gauging the CFT by  parity symmetries (preserving
conformal invariance)  of the two-dimensional space-time one obtains
a CFT on non-oriented surfaces, such as the Klein bottle. In this
case, there are physical states called crosscap states associated
with the parity symmetries  that  contain the information on the
non-orientability of the space-time. There is one crosscap state for
each parity symmetry. Such theories are usually referred to as
non-oriented Conformal Field Theories and have a lot of  interest in
string phenomenology \cite{Ibanez:2012zz} and recently in modern
advances in condensed matter physics
\cite{Cho:2015ega,Chan:2015nea,Hsieh:2014lba}.

In this work, we explore the relation between the entropies
described above in some oriented unitary and non-unitary CFT's  and
in unitary non-oriented. Our motivation comes from the fact that the
LREE of boundary states as well as the boundary entropy are
determined by the boundary coefficients.  Then, the immediate
question here is if the LREE can detect the ground state degeneracy
at a boundary. Also, we want to understand how susceptible is this
degeneracy when one changes the topology of the 1+1 space-time when
going from the cylinder to the Klein bottle. To address this work we
first analyze the relation between the boundary and topological
entropies in unitary oriented CFT. An analog to the boundary entropy
in non-oriented theories has been given in
\cite{Tu:2017wks,Tang:2017xjc}. There, the ``boundary entropy'' in
non-oriented surfaces is defined in terms of the quantum dimensions
$d$ and ${\cal D}$ of the theory, and their results were obtained
from the analysis of the Klein bottle partition function in the
open-channel. However, from their results,  a relation between their
entropy and a topological entropy is not evident. This leads us to
study the ``boundary entropy''  in non-oriented theories in the
closed-channel. By the open-closed duality, we give an equivalent
definition of this entropy in terms of the crosscap coefficients,
hence we call it the crosscap entropy.  However, compared to the
boundary states, the crosscap coefficients are expressed in terms of
both the modular transformation $P$ and $S$ matrix, consequently the
crosscap entropy is not explicitly related to the topological
entropy. Therefore, based on the behavior of the character
$\hat{\chi}$ of the M\"obius strip at the long-distance limit, we
propose two new quantities $\hat d$ and $\hD$ given in terms of the
of the modular transformation $P$ between the open and closed
channel of the M\"obius partition function. In a similar way to the
oriented case, we define ${\hat S} = \ln P_{0i}= \ln(\hat d/\hat
D)$.  With this new entropy, the crosscap entropy can bee regarded
as a difference of entropies, very similar to the boundary entropy.
Finally, we compute the LREE for 1+1 non-oriented CFT and we see
that this entanglement entropy codifies the crosscap as well as as
subleading terms.

 The organization of the article is as follows. In section
\ref{sec:REV}, we give a short review of boundary and crosscap
states in oriented and non-oriented surfaces. The definitions of
boundary and topological entropies are also outlined in this
section. In section \ref{sec:rlationBEaTE}, a relation between the
boundary and the topological entropies is described for the oriented
case. It is found that the boundary entropy is related to the
difference between the topological entropy for the non-trivial
representation and one-half of the topological entropy for the
identity representation.  In the non-oriented case, a similar
formula for the crosscap entropy is given. We compute the boundary
and crosscap entropies using our new definitions for three minimal
models of A-A series: the critical Ising model, the Tricritical
Ising model, and Tetracritical Ising model. Our results agree with
those using explicitly the form of the boundary and crosscap states.
Section \ref{sec:LREE}, is devoted to computing the LREE for
crosscap states. We also consider crosscap entropy and the LREE for
$SU(2)_k$ WZW models. In the end, the final remarks are given in
section \ref{sec:final}.

\section{Overview of entropies in Rational Conformal Field Theory}
\label{sec:REV}

In this section, we start with a  brief overview of the
two-dimensional Rational Conformal Field Theory (RCFT). We focus
basically on two-dimensional conformal field theories on surfaces
with boundaries and crosscaps. (For an extensive study of the
subject see
\cite{Fuchs:1997af,Recknagel:2013uja,Blumenhagen:2009zz,Brunner:2002em}).
We then revisit the definitions of boundary and topological
entropies and apply these definitions to some minimal models.

\subsection{Boundaries and Crosscaps in RCFT}

A  RCFT is a realization of a symmetry algebra ${\cal A}\otimes
{\bar{\cal A}}$  where the number of primary fields is finite. The
holomorphic (chiral) algebra ${\cal A}$  is an extension of the
Virasoro symmetry algebra. The generators are denoted as ${ W}
^{(r)}$ with  $r= 0, 1, 2, \ldots$ labelling the different chiral
fields with the case $r=0$ representing the Virasoro generators $L_n
= W^{(0)}$. The antiholomorphic (anti-chiral) algebra $\bar{\cal A}$
is generated by  ${\overline W} ^{(r)}$. In the following the
isomorphism between ${\cal A}$  and ${\bar{\cal  A}}$ is assumed.
The Hilbert space is organized into irreducible representations of
this symmetry algebra, where each irreducible representation  ${\cal
H}_i$  of ${\cal A}$ consists of a  tower of states constructed from
a primary field $\phi_i$ of conformal weight $h_i$ and their
descendants. Here we will denote the identity representation by $0$.

The presence of boundaries in the theory breaks the two-dimensional
conformal algebra ${\cal A}\otimes {\bar{\cal A}}$. However, one can
introduce them in such a way that a diagonal subalgebra ${\cal
A}_{diag}$ is preserved. Thus, the boundary conditions can be
described by certain coherent states called boundary states, where
all information of the boundaries is codified.  For each boundary
$a$ there is a boundary state defined as

\be \label{eq:boundstate} |B_a \rangle = \sum_i B_{a i} \, \vert {i}
\rra, \ee where $\vert {i}\rra  $ is the Ishibashi state of the
representation $i$.

On the other hand, non-oriented theories are constructed by gauging
the oriented theory with a  parity symmetry. The  simplest case is
the standard  operator $\Omega$ which acts on the spatial
coordinates as an inversion and it acts on the representations as
$\Omega: i \rightarrow \bar i$. Moreover, if the theory contains an
internal symmetry $G$, the parity symmetry operator is defined as $
\P=\Omega h$ with $h$ an internal transformation.  These  theories
are defined on non-oriented surfaces which naturally include
crosscaps.  The crosscap boundary conditions are  implemented in the
theory as coherent states called crosscap states, and for each
parity symmetry operator $\mathscr{P}_\mu$ with ${\P}_\mu ^2 = 1$
($\mu$ labels the different  parity symmetries) there is a
crosscap state defined as \be \label{eq:crosstate} |C_\mu \rangle =
\sum_i \Gamma_{\mu i} \vert C_i \rangle\!\rangle\,, \ee with $ \vert
C_i \rangle\!\rangle$ the Ishibashi crosscap state
\cite{Ishibashi:1988kg}. The coefficients $B_{a i}$ in
\eqref{eq:boundstate} and  $\Gamma_{\mu i}$ in \eqref{eq:crosstate}
are determined by the sewing constrains
\cite{Lewellen:1991tb,Fioravanti:1993hf}. Nevertheless, solving the
constrains requires a detailed analysis of the $n$-point functions
in the presence of boundaries and crosscaps. Only in few cases the
constrains  can be solved.

There is another constraint for the boundary and  the crosscap
coefficients that can be used to determine these coefficients. This
is the duality open-closed (direct-transverse) channel for the
partition functions on the annulus $\cal C$, the M\"{o}bius strip
$\cal M$ and the Klein bottle $\cal K$. This duality establishes
that, for each surface, the partition function $Z$ defined in the
open channel is the same as the partition function described in the
closed channel. The fundamental region for these surfaces is a
strip. In the open (direct) channel, the width $L$ of the strip lies
on the  spacial direction and the height lies along the temporal
direction with period $\beta$. The open-closed duality interchanges
spatial and temporal directions, therefore, in the closed
(transverse) channel the width $L$ runs along a temporal direction
and the height $\beta$ is periodic along the spacial direction. This
duality is very well known and we only mention the relations between
the partition functions following the notation given in
\cite{Brunner:2002em}:
\begin{align}
\label{eq:ocpf}
Z^{\cal C}  := \quad &\,\text{Tr}_{{\cal H}_{a b}}\, e^{-\beta H_o(L)}  =\langle B_a |e^{-L H_c(\beta)}|B_{b}\rangle \,,\nonumber \\
Z^{\cal K}:= \quad &\,\text{Tr}_{{\cal H}_{g_{\tiny{
                \mu \nu}}}}{\P}_{ \nu}\, e^{-\beta H_c(L)} =
\langle{C_{\mu}}|  e^{-\frac{L}{2} H_{c}(2 \beta)} |{C_\nu}\rangle \,,\\
Z^{\cal M}:= \quad \,&\text{Tr}_{{\cal H}_{a \mu(a)}}\,\P_\mu
e^{-\beta H_o(L) }  =\langle  B_a |e^{-\frac{L}{2} H_c(2 \beta)}
|C_{\mu} \rangle \,. \nonumber
\end{align}

The left-hand side of these equations represent the partition
functions $Z$ on ${\cal C}$,  ${\cal K}$ and  ${\cal M}$ in the open
channel,  respectively. The traces are on the Hilbert space with respective boundary and/or crosscap
conditions: ${\cal H}_{a b}$ is the Hilbert space with boundary conditions  $a$ and $b$. ${\cal H}_{g_{\mu \nu}}$  is the space of states with boundary conditions twisted  by ${g} \in G$ and by the parity symmetries $\P_\mu$. When transforming  to the closed channel,  $g= \P_\mu (\P_\nu)^{-1}$  with  $\P_\mu$ and $\P_{\nu}$ determining the crosscap boundary conditions. Finally, ${\cal H}_{a \mu(a)}$ denotes the space of states with a boundary $a$ and the image $\mu(a)$ of it under  $\P_{\mu}$.On the right-hand side  $Z^{\cal C}$,  $Z^{\cal K}$ and  $Z^{\cal M}$ denote the propagator between boundary states, crosscap states, and a boundary and a crosscap state respectively.

The equalities in  \eqref{eq:ocpf} are established by modular
transformations. In terms of the characters,    $Z^{\cal C}$ is
expressed as  $\sum_{i}n^i_{ab} \chi_i(\tau) = \sum_i B^*_{a i} B_{b
i} \chi_i(-1/\tau)$  where  $ \chi$ satisfies the modular
transformation S: $\chi_j( -1/\tau) = \sum_i \chi_i(\tau) S_{ij}$
     and the coefficients in the open channel $n^i_{ab}$ are non-negative integers.  For the Klein bottle  $Z^{\cal K}$  one has  $ \sum_i k^i_{\mu \nu} \chi(2\tau) = \sum_i \Gamma_{\mu i}^* \Gamma_{\nu i} \chi_i(-1/2\tau)$. The relation between both sides is determined by the $T$ modular transformation  $\chi_i(\tau +1)= T_{ij}\chi_j(\tau)$  followed by an $S$ transformation.  The open channel coefficient  $k^i_{\mu \nu}$  has to be an  integer satisfying
     \be
     \label{eq:intconklein}
      |k^i_{\mu \nu}| \leq h^{ii}_{\mu \nu}\quad  \text{and}  \quad  k^i_{\mu \nu} \equiv h^{ii}_{\mu \nu} \quad  (\text{mod}\,\,2)\,,
      \ee
       where $h_{ij}$ are the multiplicities of ${\cal H}_i \otimes {\cal H}_j$ in the space ${\cal H}_g$ of $g$-twisted closed string states. This condition projects out states from the oriented one. The consistency of the theory with  boundaries and crosscaps is determined by  $Z^{\cal M}$. The relation  \eqref{eq:ocpf} in this case is $ \sum_i  m^i_{a \mu} \hat{\chi}(\tau) = \sum_i B^*_{a i} \Gamma_{\mu i} \hat{\chi}_i(-1/4 \tau)$ with $\hat{\chi}(\tau) =
     e^{-\pi i (h-c/24)} \chi(\tau + 1/2)$. It is transformed as   $\hat{\chi}_j(-1/4\tau) = \sum_i \hat{\chi}_i(\tau) P_{ij}$ by the modular transformation  $P =\sqrt{T}\,S\,T^2\,S\,\sqrt{T}$. The   open-channel M\"{o}bius partition function is interpreted as the projection of states in the annulus $Z^{\cal C}$. This is achieved requiring  the coefficient $m^i$ to be an integer such that

     \be
     \label{eq:intconmob}
      m^i_{a \mu} \leq |n^i_{a \mu(a)}|  \quad \quad  \mbox{ and}  \quad \quad m^i_{a \mu} \equiv n^i_{a \mu(a)} \, \text{mod}\, 2.
      \ee
After the modular transformations described above, the open-closed
duality set  the following  constraints

\begin{align}
    \label{eq:constraints}
    Z^{\cal C}  : \quad &\, n^i_{ab} = \sum_j B^*_{a i} B_{b i} S_{ij}\, ,\nonumber\\
    Z^{\cal K}: \quad &\,  k^i_{\mu \nu} = \sum_j \Gamma_{\mu i}^* \Gamma_{\nu i}S_{ij}\,, \\
    Z^{\cal M}: \quad \,& m^i_{a \mu} = \sum_j B^*_{a i} \Gamma_{\mu i} P_{ij}  \,.\nonumber
\end{align}

For conformal field theories with isomorphic left- and
right-algebras some solutions to these constraints have been found.
We will consider the solution in which the internal symmetry is the
identity, and the  multiplicity  $h^{ii}$  is a charge conjugation
invariant. In such case the boundary labels $a$ take the  same
values of the representations $i$, therefore  one can set $n^i_{ab}
= N^{\overline{i}}_{ab}$ with  $N^{\overline{i}}_{ab}$ the fusion
coefficients in the Verlinde formula. In this way, the solution to
the first equation in \eqref{eq:constraints}  is
  \be
  \label{eq:carsol}
  B_{ai}= \frac{S_{ai}}{\sqrt{S_{0i}}} \,,
  \ee
 This is known as the Cardy solution \cite{Cardy:1989ir}.
With the same considerations, the relevant parity operator is the
standard  operator $\Omega$ for which the index $\mu$ agrees with
the index of the identity representation   $(\mu = 0)$. The
solutions to the last two  equations in \eqref{eq:constraints} are
found using the integrality conditions \eqref{eq:intconklein} and
\eqref{eq:intconmob}. In \cite{Pradisi:1995qy},
Pradisi-Sagnotti-Stanev (PSS) realized that the matrices $S$ and $P$
satisfy a similar relation to the Verlinde formula: $Y_{ij}^k =
\sum_{\ell} \frac{S_{i \ell} P_{j\ell} P_{k\ell}^*}{S_{0 \ell}}$ and
that the solutions to \eqref{eq:intconklein} can be written in terms
of this matrix as
\begin{align}
m^j_{i0} &= Y_{\overline{i}0}^{\overline{j}} , \nonumber\\
k^i_{00}& = Y^0_{i0}.
\label{newsolutions}
\end{align}
Using these expressions  in  equation  \eqref{eq:constraints}, the PPS solution for the  crosscap
coefficients is given by

\be
\label{eq:psssol}
\Gamma_{0 i}= \frac{P_{0i}}{\sqrt{S_{0i}}}.
\ee

If the theory has an internal symmetry associated to simple currents
$J$ of order $N$ the crosscap coefficients are given by
\cite{Huiszoon:1999xq}
\be \label{eq:simcurcoe} \Gamma_{[J^n]i} =
\frac{P_{J^n i}}{\sqrt{S_{0i}}},  \quad \quad \quad n=0,\ldots,N-1,
\ee
with the condition (\ref{newsolutions}) modified by replacing $0$ by the index running on the algebra representations.
\subsection{Boundary and Topological entropy}

The topological entropy is a measure of topological order manifested
as  entanglement in the ground state of a system
\cite{Kitaev:2005dm,Levin:2006zz}. It is defined as \be
\label{eq:topent} S^{{top}}_i =  \text{ln} \Bigl({\frac{d_i}{\cal
D}}\Bigr) = \ln (S_{0i}) \,, \ee where $d_i$ is the relative quantum
dimension of the representation ${\cal H}_i$  with respect to the
identity representation ${\cal H}_0$ and ${\cal D} = \sqrt{\sum_i
d^2_i}$ is the total quantum dimension
\cite{Dijkgraaf:1988tf,Moore:1988qv}.   These quantum numbers are
related to the matrix $S$  as $ d_i=S_{0i} /{S_{00}}$ with $\D =
1/S_{00}$.

On the other hand, the boundary entropy is associated to the ground
state degeneracy due to the boundaries. It can be computed from the
partition function on the cylinder with boundary conditions $a$ and
$b$. In the closed channel  and in the thermodynamic limit $L \gg
\beta$, the ground state degeneracy is $g=g_a g_b$. The
contribution of the boundary $a$ to the ground state degeneracy is
$g_a = \langle 0|B_a\rangle=B_{a 0}$ with $|0 \rangle$ the
ground state of  $H_c$. Due to the open-closed duality, the ground
state degeneracy can also be computed in the loop-channel, it is
given as $ g_a g_b =\sum_i n^i_{ab} S_{0i}$, where the coefficient
$n^i_{ab}$ are the non-negative integers described above and $S_{0i}$ a non-negative
number \cite{Affleck:1991tk}.

The boundary entropy  associated to the ground state degeneracy $g_a$ is defined as
\be
\label{eq:boundent}
S^B_a = \text{ln}\, g_a  \quad \text{with} \quad g_a= \begin{cases}
\sqrt{\sum_i n^i_{aa}\,S_{0i}} & \text{open channel}\\
B_{a0}& \text{closed channel}\,.
    \end{cases}
\ee

The ground state degeneracy is not only related to the boundary
entropy, it was observed in  \cite{Harvey:1999gq} that the
regularized dimension of ${\cal H}_{aa} =  \oplus_i n^i_{a a} {\cal
H}_i$  is precisely the ground state degeneracy \be
\label{eq:regdimcylinder} \text{dim} {\cal H}_{aa} =
\lim_{q_o\rightarrow 1} q_c^{c/24} Z^{\cal C}_{aa} (q_o) = g^2_{a} =
\sum_i n^i_{aa} \frac{d_i}{\D}\,, \ee here $Z^{\cal C}_{aa} (q_o)$
is the partition function in the open channel described in
\eqref{eq:ocpf}.

\section{Relation between Boundary entropy and Topological entropy}
\label{sec:rlationBEaTE}

So far we have discussed the boundary entropy and topological
entropy separately. In this section we study the relation between
them  for some unitary minimal  models, specifically in some of the
A-A series of the minimal models.
\subsection{Oriented case} \label{subsec:oriented}
The relation between the boundary entropy  and the topological
entropy  comes from the following observation. A boundary state is
an entangled system in its left- and right modes
\cite{PandoZayas:2014wsa,Das:2015oha}, therefore one could expect
that  the boundary entropy defined by the coefficients of the
boundary states, must be associated to some  measure of entanglement
of  the ground state. In the closed channel, the boundary entropy is
given as $S^B_a = \ln\,(B_{a0})$. Using the definition
\eqref{eq:carsol} for $B_{a0}$, we have  $S^B_a = \ln (
S_{a0}/\sqrt{S_{00}}) = \ln (d_a /{\D})  -\ln\sqrt{1/\D} $, where in
the last equation we have taken into account that $S$ is symmetric
and  that $S_{0i} = d_i/{\cal D}$. Therefore,  by equation
\eqref{eq:topent} we have
 \be
 \label{eq:relboutop}
 S^B_a  =  S^{top}_a -\frac{1}{2} S^{top}_0 \,.
 \ee

Note that for the identity representation   $d_0=1$ and the
topological entropy in this case is   $S^{top}_0= -\ln{\cal D} = \ln
(S_{00})$. The boundary entropy associated to the boundary $a$  can
be computed easily from the relation \eqref{eq:relboutop}, since it
is the difference between the topological entropy for the
irreducible representation-$a$ and the topological entropy of the
identity representation. In particular, $S_0^B = \frac{1}{2}
S_0^{top}$. While the topological entropy is always negative (since
$d_i < {\cal D}$), the sign of the boundary entropy  depends on the ground state
degeneracy  which is  $g_a \geq 1$ for   $d_a \geq \sqrt{\cal D}$
and $g_a < 1$ for   $d_a < \sqrt{\cal D}$, as can be inferred from
\eqref{eq:relboutop}.

In the open channel, we substitute $S_{0i}= d_i/{\cal D}$  in the
first relation of  \eqref{eq:boundent} to have $g_a = \sqrt{\sum_i
n^i_{aa} d_i/{\cal D} }$.   Using the fact that  the Cardy solution
sets $n^k_{aa} =N^k_{aa}$  and that the quantum dimension satisfies
the fusion algebra $d_i d_j = \sum_k N_{ij}^k d_k$
\cite{DiFrancesco:1997nk}, we get equation \eqref{eq:relboutop}.

We apply  equation \eqref{eq:relboutop} to some minimal models of
the A-A series where the $S$-matrix is given in
\cite{DiFrancesco:1997nk,Bianchi:1991rd}. The Ising model with
$c=1/2$ and ${\cal D}= 2$ contains three primary operators: the
identity operator $\bf{1}$ (labeled by $0$), the energy field
$\epsilon$ and the spin field $\sigma$. For the operators $\bf{1}$
and $\epsilon$ which have the same relative dimension $d=1$, the
topological entropy is $S^{top}_{{0},\epsilon} = - \,\ln\, 2$. The
associated boundary entropy is $S^B_{0,\epsilon} = -\frac{1}{2}
\text{ln}\,2$. For the spin operator the relative dimension is
$d_\sigma=\sqrt{2}$ and $S^{top}_\sigma= -\frac{1}{2} \text{ln}\,2$,
hence $S^B_{\sigma} = 0$.  Since $d_\sigma$ is the square of the
total quantum dimension the boundary entropy is zero. These results
agree with those given in \cite{Affleck:1991tk}.

In the Tricritical model with $c=7/10$ there are six primary fields
$\big( \bf{1}, \epsilon, \epsilon^\prime, \epsilon^{\prime
\prime},\sigma, \sigma^\prime \big)$, the relative dimensions are
$d_{0} = d_{\epsilon^{\prime \prime}} =1,\, d_{ \epsilon}=
d_{\epsilon^\prime} = s_1/s_2, d_\sigma=\sqrt{2} s_1/s_2$ and $d_{
\sigma^\prime}=\sqrt{2}$ and  ${\cal D}=\sqrt{40/(5-\sqrt{5})}$.

The entropies are
\begin{align}
S^{top}_{0} & = -\ln \D = -1.3361 & S^B_{0} & = \ln \sqrt{s_2}= -0.6681 \nonumber \\
S^{top}_{\epsilon} & = \ln (s_1) = -0.8549 & S^B_{\epsilon} & = \ln \Big(\frac{s_1}{\sqrt{s_2}}  \Big) = -0.1868\nonumber \\
S^{top}_\sigma &= \ln (\sqrt{2}s_1) = -0.5083 & S^B_\sigma& = \ln \Big(\sqrt{\frac{2}{s_2}}s_1 \Big) = 0.1597\nonumber \\
S^{top}_{\sigma^\prime}&  = \ln (\sqrt{2} s_2) = -0.9895 &  S^{B}_{\sigma^\prime} & = \ln \Big(\sqrt{2 s_2}  \Big) = -0.3215
\end{align}
with $s_1= \sqrt{(5+\sqrt{5})/40}$ and $s_2 =\sqrt{(5-\sqrt{5})/40} $. Also  $S_{\epsilon^{\prime \prime}} =S_{0}$ and $S_{\epsilon^{\prime}}= S_\epsilon$ since such fields have the same relative quantum dimension, respectively. There is not a $d_i$ equal to $\sqrt{\D}$, so all boundary entropies are different from zero.
There is a positive boundary entropy corresponding to the primary field $\sigma$, since  its relative dimension satisfies $d_\sigma > \sqrt{\D}$.  These expressions for the boundary entropies are in complete agreement with those computed directly from the boundary states in this model.

In the Tetracritical Ising Model with $c = 4/5$ the primary fields
have conformal weight $\{0,2/5,1/40,7/5, 21/40,
1/15,3,13/8,2/3,1/8\}$ and $\D = \sqrt{6(5+\sqrt{5})}$. As in the
Tricritical case all boundary entropies are different from zero, the
positive boundary entropies correspond to  the operators with
conformal weight $7/5, 1/40$ and $21/40$. All other boundary
entropies are negative.

\subsection{Non-unitary minimal models}

In this subsection we study the relation between the boundary
entropy and the topological entropy in non-unitary Virasoro minimal
models.

These theories contain fields with negative conformal weights. The
ground state in these theories is not the conformal vacuum  but the
state with lowest conformal weight. The   representation of the
field with smallest conformal weight is denoted as $\m$.  When
boundary conditions are included, the boundary coefficients for the
A-series are \cite{Runkel:1998he}: \be \label{eq:bscnonuni} g_a = B_{a\, \m} =
\frac{S_{a\, \m}}{\sqrt{S_{0 \,\m}}}\,. \ee

On the other hand, a {\it generalized quantum dimension} for these
models was defined in \cite{Nahm:1992sx, Terhoeven:1993sn,
Terhoeven:1995jz} as: \be \label{eq:gerqd} \breve{d}_{i} = \lim_{q_o
\rightarrow 1}\frac{{\chi}_i(\tau)}{{\chi}_0(\tau)} =  \lim_{{q_c}
\rightarrow 0}\frac{ \sum_{j}S_{ji}{\chi}_j(-\frac{1}{
\tau})}{\sum_{j}S_{j0}{\chi}_j(-\frac{1}{ \tau})}  = \frac{S_{i \,
{\rm min}}}{S_{0 \, {\rm min}}} \,. \ee

The elements $S_{i \, \m }$ of the matrix S  are positive for any
$i$-representation.  We define the   {\it total generalized quantum
dimension} as $\breve{{\D}} = \sqrt{\sum_i \breve{d_i^2}}$ and since
$S$ is unitary we can express this quantity as $\breve{{\D}} =
1/S_{0 \, {\rm min}}$. The topological entropy is then defined as \be
\label{eq:topentnonuni} S^{\text{top}}_i =  \text{ln}
\Bigl(\frac{\bd_i}{\breve{\D}}\Bigr) = \ln (S_{i \, {\rm min}}) \,. \ee
Substituting the expression \eqref{eq:bscnonuni} into
\eqref{eq:boundent} and using the equation \eqref{eq:topentnonuni}
we get the relation  \eqref{eq:relboutop} between the boundary  and
the  topological entropies.

As an example we apply equation \eqref{eq:relboutop} to  the
Yang-Lee model. This is a  non-unitary minimal model of the A-series
with central charge $c=-22/5$. It has only  two primary fields
denoted as  $\bf{1}$ and $\phi$ with conformal weight $h=0$ and
$h_{\m}=-1/5$, respectively. The matrix $S$ for this model can be
found in \cite{Terhoeven:1995jz}, the relevant matrix elements for
our study are $S_{0 \,\m}= (2/\sqrt{5}) \text{sin}(\pi/5)$ and
$S_{\m \, \m}= (2/\sqrt{5}) \text{sin}(2\pi/5)$.

The quantum dimensions are $\bd_0 =1, \bd_{\m}=
\sqrt{(3+\sqrt{5})/2} = 1.61803$ and
$\breve{{\D}}=\sqrt{(3+\sqrt{5})/2} = 1.90211$. The topological and
boundary entropies are:

\begin{align}
    S^{\text{top}}_0& = -0.64297 \,, &      S^{\text{B}}_0 &= -0.32148  \nonumber\\
    S^{\text{top}}_\m &= -0.16175 \,, &     S^{\text{B}}_\m &= 0.15973
\end{align}
The results for the boundary entropy coincide with those in
\cite{Dorey:1999cj}, which were obtained computing the correlation
functions on a disk.  As before, we observe that $S^B_0 < 0$ as long
as $\bd < \sqrt{\breve{{\D}}}$ and $S^B_\m > 0$ since $\bd \geq
\sqrt{\breve{{\D}}} $.

It is worth mentioning that in the Yang-Lee model their characters can
be represented both in the product and in the sum representations.
In the last one,  they are written as \cite{Nahm:2004ch}
\begin{equation}
\chi_0^{(2,5)} = q^{11/60} \sum_{n \in \mathbb{N}} {q^{n(n+1)} \over
(q)_n}, \ \ \ \ \   \chi_1^{(2,5)} = q^{-1/60} \sum_{n \in
\mathbb{N}} {q^{n^2} \over (q)_n},
\end{equation}
where $(q)_n = (1-q)(1-q^2) \cdots (1-q^n)$. A more general
structure of these characters is given by the so called Nahm's sum
\begin{equation}
 \chi_{A,B,C}(\tau) = \sum_M a_M q^M = \sum_{{\bf n} \in \mathbb{Z}^r} {q^{{\bf n}^t A {\bf n} + B \cdot {\bf n} +C}
 \over
 (q)_{\bf n}},
\label{nahmSum}
\end{equation}
where $(q)_{\bf n} = (q)_{n_1} \cdots (q)_{n_r}$ and $(q)_{n_1}
=\prod_{I=1}^{n_r} (1 - q^I)$ with $q=e^{2\pi i \tau}$ and ${\bf n}
=(n_1, \dots, n_r)$. Moreover $A$ is a $r \times r$ symmetric and
positive matrix, $B$ is a vector and $B \cdot {\bf n}= \sum_{I=1}^r
b_I n_I$ and  $C$ is an scalar. The specific form of the matrix $A$, the vector $B$ and the
scalar $C$ is related to a certain CFT. For instance, for the
Yang-Lee model, their characters can be obtained from this general
form (\ref{nahmSum}) for specific values of $r$, $A$, $B$ and $C$.

According to Nahm's conjecture \cite{Nahm:2004ch}, the characters
(\ref{nahmSum}) are modular forms for certain values of $A$, $b$ and
$C$, the entries of the matrix $A$ satisfy certain equations
associated to a certain integrable model. Two dimensional integrable
models described, for instance, by a massive scalar quantum field
theory can be regarded as a CFT at high energies, i.e. in the limit
when the mass of the field can be considered very small compared
with the characteristic energy of the relevant process. Thus
integrable models can be interpreted as some deformations of CFTs
preserving certain aspects of the theory, for instance, the infinite
hierarchy of conserved quantities. In these integrable models the
Thermodynamic Bethe Ansatz (TBA) constitutes a bridge between
integrable models and CFTs
\cite{Nahm:1992sx,Terhoeven:1993sn,Nahm:2004ch}. Some properties of
the CFT as the conformal weights and the central charge can be
written in terms of some functions as the dilogarithm formula
\cite{Nahm:1992sx,Terhoeven:1993sn}.

In Ref.  \cite{Terhoeven:1993sn} it was shown that after an
asymptotic treatment of a continuous version of Eq. (\ref{nahmSum})
and the use of the saddle point method, the TBA-type equation
realization of the partial generalized quantum dimension (mentioned
above) is given by \cite{Terhoeven:1993sn}
\begin{equation}
\breve{\lambda}_{i} \equiv {S_{i,{\rm min}} \over S_{0,{\rm min}}}=
\prod_{I=1}^r\bigg(1 -w_I\bigg)^{(b_{I,i} -b_{I,0})},
\end{equation}
where
\begin{equation}
S_{i,{\rm min}} = \exp \bigg\{\sum_I \bigg(b_{I,i} -{1 \over
2}\bigg) \log (1 -w_I) \bigg\},
\end{equation}
with $w_i$ are real parameters with $0<w_I<1$ for all $I=1,\dots,r$.
This is an important point for our description since at least for
these classes of systems is possible to define a corresponding
analogue of the topological entropy
\begin{equation}
{\cal S}^{top}_i = \ln (S_{i,{\rm min}})= \sum_{I=1}^r \bigg(b_{I,i}
-{1 \over 2} \bigg) \ln (1-w_I),
\end{equation}
which evidently constitutes a finite value.

\subsection{Non-oriented case}

In non-oriented theories, the presence of parity symmetries modifies
the space of states since the states can have positive or negative
parity eigenvalues. For the Klein bottle surface, the physical
states are those which are symmetric under parity symmetries. In
particular, if the ground state is degenerated, the antisymmetric
part is projected out  \cite{Stanev:2001na}. Therefore one could
expect an entropy associated to such degeneracy. In
\cite{Tu:2017wks} the {\it boundary entropy} associated to the
standard parity projection was defined using $Z^{\cal K}$ in the
open channel. There, the ground state degeneracy is given as $g =
\sum_i k^{i}\, d_i/\D$ were $k^i$  are the non-negative coefficients
of the Klein bottle partition function in the open channel subjects
to the duality constraints. We  want to compute this entropy in the
closed channel where we have the crosscap state formalism at hand.
In the next, we prefer to call the entropy due to the presence of
crosscaps, the {\it crosscap entropy}.

To obtain the  crosscap entropy, we follow the prescription  given
for the boundary states in \cite{Affleck:1991tk}. The Klein bottle
amplitude in  the transverse channel is given as
\cite{Brunner:2002em}
    \be
    Z^{\cal K} = \langle C_\mu |e^{-\frac{\pi L}{2\beta} (L_0 + \bar L_0 - \frac{c+\bar c}{24})} | C_\nu \rangle = \sum_i \Gamma_{\mu i}^*  \Gamma_{\nu i}\, \chi_i (-1/{2\tau})\, .
    \ee
    In the limit $L \gg\beta$, only the ground state dominates and
        $
    \label{eq:pfKgsc}
    Z^{\cal K}  \sim  \langle C_\mu | 0\rangle \langle 0 |C_\nu \rangle  \, e^{\frac{\pi c L}{24 \beta}} =  \Gamma^*_{\mu 0} \Gamma_{\nu 0}\, e^{\frac{\pi c L}{24 \beta}}\,,
    $
    where $\langle 0 |C_\mu \rangle= \Gamma_{\mu 0}$ with $|0 \rangle$ the ground state  of $H_c$.
    The thermodynamical entropy is
    $
    S_{\cal K}^{ther} = \frac{cL\pi}{12 \beta} + \ln g_{\mu} +  \ln g_{\nu} \,,
    $
where   $g_\mu= \Gamma_{\mu 0}$ is the  degeneracy of the ground
states  due to the crosscap associated to the parity symmetry ${\cal
P}_\mu$. Summarizing, we have that the crosscap entropy is

    \be
    \label{eq:crosent} S^C_\mu = \text{ln} \, g_\mu \quad \quad \text{with} \quad \quad  g_\mu = \begin{cases}
        \sqrt{\sum_i k^i_{\mu \nu}\,\frac{d_i}{D}} & \text{open channel}\\
        \Gamma_{\mu 0}& \text{closed channel}\,.
    \end{cases}
        \ee

Now we want to address if this crosscap entropy can be expressed as
a difference of topological entropies as in equation
\eqref{eq:relboutop}. In the open channel,  the ground state
degeneracy  defined in \eqref{eq:crosent} is given in terms of the
quantum dimensions but it is not evident how it can be expressed in
terms of topological entropies. On the other hand, we are not aware
if from the topological quantum field theory point of view, the
topological entanglement entropy \eqref{eq:topent} can be
consistently defined for no-oriented surfaces, therefore we can not
define the crosscap entropy as a difference of topological entropies
as in \eqref{eq:relboutop}.

We know from equations \eqref{eq:psssol} and \eqref{eq:simcurcoe}
that $\Gamma_{\mu i} = P_{\mu i}/\sqrt{S_{0i}}$. We recall that the
label $\mu$ denotes the kind of parity operators and  for each
operator there is a crosscap entropy. For  the  standard parity
operator the index $\mu$ agrees with the index of the identity
representation ($\mu = 0$), and for non-standard parities associated
to simple currents $\mu$ takes the label of the simple currents.
Then, the crosscap entropy in the closed channel is $S_\mu^C = \ln
\Gamma_{\mu 0} = \ln (P_{\mu 0}) - \ln(\sqrt{S_{00}})$.  Now, we
define the quantity

\be
\label{eq:nontopent}
{\hat S}_{i} =  \text{ln} P_{i0} =  \text{ln} \Bigg( \frac{{\hat d}_i}{\hat{{\cal D}}}\Bigg)    \quad \quad \text{for} \quad  P_{0i} \neq 0.
\ee
This is defined for each irreducible representation in which  $P_{ 0i}$ is not zero since primary fields $\phi_i$ which are not invariant under the parity symmetry  have $P_{0i}=0$.

The term $P_{0i}$ appears naturally in the M\"obius  partition function $Z^{\cal M}$. In terms of this,  the regularized dimension of ${\cal H}_{a \mu(a)}$ is
\begin{align}
\text{dim} {\cal H}_{a \mu(a)} &= \lim_{q_o\rightarrow 1} q_c^{c/96} Z^{\cal M}(q_o) =   \lim_{q_o\rightarrow 1} q_c^{c/96} \,\Tr_{{\cal H}_{a \mu}}\P_\mu e^{2\pi i \tau H_o}\nonumber \\
&  =\lim_{q_o\rightarrow 1} q_c^{c/96} \,\sum_i m^i_{a \mu(a)}\hat{\chi}_i(\tau)  \\
&= \sum_i m^i_{a \mu(a)} P_{0i}\,,\nonumber
\end{align}
where in the first line, $H_o$ is the Hamiltonian in the open
sector. In the second row $m^i_{a \mu(a)}$ are the integers whose
values are restricted by the open-closed duality. Here,
$\hat{\chi}_i(\tau)$ is defined up to a factor of $1/2$ which is
induced by the relative orientation of the horizontal sides of the
cylinder with a boundary and a crosscap and the ends. Due to this
$1/2$, the  signs of the states alternate in  the character  and the
signs depend on the levels. In the last equality we have used the
modular transformation  $\hat{\chi}_i(-1/4\tau) = \sum_i
\hat{\chi}_i(\tau) P_{ij}$. From this modular transformation  the
regularized dimension is expressed in terms of the elements of the
matrix $P$ and therefore it can not be expressed in terms of the
quantum dimensions $d$ and $\cal D$.

Looking at the character of the M\"obius partition functions  $\hat
\chi_i$,  we note that in the limit  ${q \rightarrow 1}$, the
contributions to $ \hat{\chi}_i(\tau)$ at each level are the
differences between the number of  even and odd states under the
parity operator. Hence, we define a relative {\it parity} index of
the representation $ i$ compared to the identity representation $ 0$
as \be \label{eq:relparind} {\hat d}_i = \lim_{q_o \rightarrow
1}\frac{\hat{\chi}_i(\tau)}{\hat{\chi}_0(\tau)} =  \lim_{{q_c}
\rightarrow 0}\frac{ \sum_{j}P_{ji}\hat{\chi}_j(-\frac{1}{4
\tau})}{\sum_{j}P_{j0}\hat{\chi}_j(-\frac{1}{4 \tau})}  =
\frac{P_{0i}}{P_{00}} \,. \ee For the minimal models here analyzed,
$P_{0i} \geq 0$ and $P_{00}>0$.  Since $P^\dagger P =  1$, then
$\sum_{i}|P_{0i}|^2 =1$ and one can define a  total parity index as
$ \label{eq:totparind} \hat{{\cal D}}= \sqrt{\sum_i {{\hat d}_i}^2 }
= \frac{1}{P_{00}}\,. $ In terms of these numbers the regularized
dimension becomes \be\label{eq:regdimmoebius}
 \text{dim} {\cal H}_{a \mu(a)} = \sum_i m^i_{a \mu} \frac{\hat{d}_i}{\hD}.
\ee
This expression is very similar to the regularized dimension in the cylinder \eqref{eq:regdimcylinder}.

With the definition \eqref{eq:topent} and \eqref{eq:nontopent} we can write the crosscap entropy as
\be
\label{eq:crosscapent}
S^C_\mu  = \hat{S}_\mu   - \frac{1}{2} S_0^{top}  \,.
\ee
 In this way the  crosscap entropy  has the same form as the boundary entropy given in \eqref{eq:relboutop}. For the standard parity, $\hat{S}_0 = - \ln\, \hat \D= \ln\, P_{00}$.

From the relation \eqref{eq:crosscapent} we can see that  $S_\mu^C > 0$ if ${\hat d}_\mu > \hD /\sqrt{\D}$.
 In some cases the equality to zero is possible \cite{Tu:2017wks}, however  for the minimal models here studied with standard parity, this inequality is true. This is because  $\hD$ and $\D$ are related to $P_{00}$ and $S_{00}$ respectively, and  $\sqrt{S_{00}} < P_{00}$. We have tested this last inequality for several values of the parameter $m$ of the minimal models in the A-A series and it is satisfied, therefore we can assert that for such models the  crosscap ground state degeneracy is $g>1$. We will see in the next chapter that in $SU(2)$ WZW models it is possible to have models  where the crosscap ground state degeneracy $g<1$.

We have computed the crosscap entropies for the same models studied
in \ref{subsec:oriented} using the definition
\eqref{eq:crosscapent}. For the Ising and the Tricritical model
(with real representations) there is only one Klein bottle partition
function associated to the standard parity symmetry (where $\mu =
0$), then  the crosscap coefficients are given by the PSS solution
and the crosscap entropy is determined by the coefficient
$\Gamma_{00} = P_{00}/\sqrt{S_{00}}$. We summarize the results in
Table \ref{tab:models}.

 \begin{table}[h!]
    \begin{center}
     \begin{tabular}{|c|c|c|c|}
    \hline
    & $\S_0 =- \ln(\hD)$ & $S_0^{top}=-\ln(\D)$ & $S_0^C$ \\
    \hline
    Ising   & $ - \ln (2/\sqrt{2+\sqrt{2}})$ & $-\ln(2)$ & $\ln \sqrt{(2+\sqrt{2})/2}$ \\
            & =$ -0.07917$ & =$ -0.69315$ &$ =0.2673$ \\
    \hline
    Tricritical  & $\ln (\sqrt{2+\sqrt{2}})s_1$ & $-\ln(s_2)$ & $\ln (\sqrt{2+\sqrt{2}}s_1/\sqrt{s_2})$ \\
                & $= -0.24093$ & $= -1.33611$ &$ = 0.4271$ \\
    \hline
        Tetracritical  & $\ln[(1+\sqrt{3})s_1/\sqrt{3}]$ & $-\ln(\sqrt{6(5+\sqrt{5})})$ & $\ln\big(\frac{1+\sqrt{3}}{(3)^{1/4}} \frac{s_1}{\sqrt{s_2}}\big)$ \\
    & $= -0.3992$ & $= -1.8854$ &$ = 0.54355$ \\
    \hline
     \end{tabular}
   \caption{For the Ising model  $\hD = 2/ (\sqrt{2 +\sqrt{2}})$ and $\D=2$. In the Tricritical model, $\hD = (\sqrt{2+\sqrt{2}}s_1)^{-1}$ and $\D = 1/s_2$. In the Tetracritical model  $\hD = \frac{\sqrt{3}}{(1+\sqrt{3})s_1}$ and $\D = \sqrt{6(5+\sqrt{5})}$ where $s_1 = \sqrt{\frac{5+\sqrt{5}}{40}}$ and  $s_2 = \sqrt{\frac{5-\sqrt{5}}{40}}$. }
    \label{tab:models}
 \end{center}
 \end{table}
 We could  also compute the crosscap entropy using the crosscap states. In such case, the crosscap coefficients for the Ising model were computed in \cite{Fioravanti:1993hf} and  the  crosscap entropy for the Ising model here computed  agrees with the given in \cite{Tu:2017wks, Tang:2017xjc}, where it was obtained working in the open channel. For the Tricritical model, only the rate between the coefficients was given  in \cite{Tsiares:16}. We have computed these coefficients in the Tricritical and Tetracritical model and we give the explicit expressions in  Appendix \ref{ap:coefficients}. In both cases, the crosscap entropy coincide with those given in Table \ref{tab:models}, respectively.


Finally, we go back to  the M\"obius partition function in the
closed sector  which in the limit $L \gg \beta$ becomes  $ Z^{\cal
M}_{a \mu(a)} \sim  B_{a0}  \Gamma_{\mu 0}\, \text{exp}(\pi Lc/24
\beta) $. As in the cylinder and Klein bottle case, we can  define
the thermodynamic entropy in the M\"obius sector, and the
contribution to this entropy due to the boundary state $|B_a\rangle$
and to the crosscap state $|C_{\mu(a)} \rangle$ is given by $S^{\cal
M} = \ln(B_{a0}) + \ln(\Gamma_{\mu 0}) = S^B_a + S^C_\mu$. This
equation agrees with the one given in \cite{Chen:2017coh} up to a
factor of $1/2$, this is because  the authors considered only the
standard parity operator and define the ground states degeneracy as
$g^2$. In our case we allow the possibility of other parity
operators like in theories with simple currents and then the ground
state degeneracy will have the form $g_\mu g_\nu$.


\section{Left-right entanglement entropy for crosscaps in RCFT}
\label{sec:LREE}
The methodology to compute the LREE for crosscaps is very similar to the
LREE for boundary states given in \cite{Das:2015oha}, here we
rewrite those computations adapted to crosscaps. In the following
discussion,  the label $\mu$ denoting different crosscap states
related to the different parity operators is intentionally omitted. In
order to define the density matrix,  one has to consider instead the
regularized crosscap state  $e^{- \frac{\varepsilon}{2} H(2 \ell)}
|C \rangle$. This is the same prescription employed in the case of
entanglement entropy for boundary states due to the fact that the
boundary Ishibashi states are non-normalizable. The parameter
$\varepsilon$ is an UV cutoff and  $H(2 \ell) = \frac{\pi}{\ell} (
L_0 + \bar L_0 - c/12)$ is the Hamiltonian defined on a circle of
circumference $2 \ell$ that generates time translations in the
tree-channel. In this configuration $\tau = i \ell/2 \varepsilon$.
The density matrix is defined as
\be
\label{eq:denmat}
 \rho = \frac{e^{- \frac{\varepsilon}{2} H}|C\rangle \langle C|
e^{-\frac{\varepsilon}{2} H}}{{\cal N}}\, , \ee
where the
normalization constant  ${\cal N} = \sum_j |\Gamma_j|^2 e^{-2 \pi
i(h_j - c/24)}\chi_j( -1/2 \tau +1)$ is fixed by the  condition
$\mbox{Tr} \rho = 1$. After performing the $T$-transformation of the
character, this normalization constant becomes ${\cal N} = \sum_j
|\Gamma_j|^2 \,\chi_j (-1/2 \tau)$.  This constant  is in fact  the
Klein bottle partition function between the same crosscap states
\cite{Blumenhagen:2009zz}.

Let us now denote the holomorphic sector by $A$ and the
anti-holomorphic sector by $B$. The reduced density matrix $\rho_A =
\mbox{Tr}_B \,\rho$, which is obtained by tracing on the
anti-holomorphic modes is given by
 \be
 \rho_A = \frac{1}{\cal N}\Tr_B ( e^{- \frac{\varepsilon}{2} H}|C\rangle \langle C| e^{-\frac{\varepsilon}{2} H} )
 \,.
 \ee
It is easy to see that this matrix is diagonal.  Using the
replica trick, and after performing the $T$-transformation on the
characters we find

\be \label{eq:Trreplica} \mbox{Tr} \rho^n_A = \frac{1}{{\cal N}^n}
\sum_j |\Gamma_j|^{2n} \chi_j (-n/2\tau)\,. \ee After this
transformation, the computation of the entanglement entropy is very
similar to the boundary states given in
\cite{PandoZayas:2014wsa,Das:2015oha}. We will not repeat those
computations here and  we give just the relevant results. One has to
perform an $S$-transformation of the characters  and take the limit
$\varepsilon/\ell \rightarrow 0$ to obtain the leading  contribution
in the characters which comes from the lowest conformal dimension.
In unitary theories, the  identity representation has the lowest
value and it is zero. Then,
\be
 \Tr_A \rho^n_A =  e^{\frac{2 \pi \ell c}{24 \varepsilon}(\frac{1}{n} - 1)}\frac{\sum_{j}
    |\Gamma_j|^{2n} S_{j0}}{\Big[\sum_{j} |\Gamma_j|^{2} S_{j0}
    \Big]^n} \,.
 \ee
Using the  definition for the entanglement entropy $S_A =-
\displaystyle{\lim_{ n \rightarrow 1}} \,\partial_n
\mbox{Tr}\rho_A^n $,  the LREE for each crosscap state associated to
${\P}_\mu$ is given by \be \label{eq:LREEcrosscap} S_{|C_{\P_\mu}
\rangle} = \frac{\pi \ell c}{6\varepsilon} - \frac{\sum_j
|\Gamma_{\mu j}|^2 S_{j 0}\, \mbox{ln}|\Gamma_{\mu j}|^2}{\sum_j
|\Gamma_{\mu j}|^2 S_{j 0}} + \mbox{ln} \sum_j |\Gamma_{\mu j}|^2
S_{j 0}\,.
 \ee
 Using the symmetric and unitary properties of the matrix $P$, the LREE becomes


\be \label{eq:collectiveEE} S_{|C_{\P_\mu}\rangle} = \frac{\pi \ell
    c}{6 \varepsilon} -  2 \Bigg( \frac{\hat{d}_\mu}{\hat{\D}} \Bigg)^2 S_\mu^C - \sum_{j \neq 0} |P_{\mu j}|^2 \mbox{ln}
\Bigg(\frac{|P_{\mu j}|^2}{S_{0j}}\Bigg).
\ee
with  $S_\mu^C$ given in \eqref{eq:crosscapent}.

For the particular case of the standard parity, we have
\be
 \label{eq:lreetop}
 S_{|C_{\P_0}\rangle}= \frac{\pi \ell c}{6\varepsilon} -\frac{2}{{\hat{\cal D}}^2} S_0^C -2 \sum_{j \neq 0} \Big(\frac{{\hat d}_j}{\hat{\cal D}}\Big)^2  \Bigl[  {\hat S}_j - \frac{1}{2}   { S}^{top}_j     \Bigr]\,.
\ee

%

\subsection{Wess-Zumino-Witten models}

The crosscap states in WZW models are also determined by the
crosscap coefficients \eqref{eq:psssol} and \eqref{eq:simcurcoe}
\cite{Pradisi:1995pp,Pradisi:1995qy,Pradisi:1996yd,Fuchs:2000cm}. In
this case the parity operator is given as  ${\cal P} = {\cal I}
\Omega$ which is a combination of $\Omega$ with a target involution
${\cal I}$ on the group manifold. For any WZW model one can consider
the involution  ${\cal I}: h \rightarrow  \gamma^n {h}^{-1}$ where
$h$ is the map from the world-sheet to the group manifold and
$\gamma$ generates the $\mathbb{Z}_n$ center of group manifold and
$n=0, \ldots, N-1$
\cite{Horava:1990ee,Huiszoon:2001wv,Brunner:2001fs}. For $n=0$, the
associated crosscap state is the PSS solution \eqref{eq:psssol}.  On
the other hand, for $n\neq 0$, the center of the group is isomorphic
to an abelian group formed  by  simple currents, therefore  the
related crosscap state is given by \eqref{eq:simcurcoe}. The matrix
$S$ for WZW model is given in \cite{DiFrancesco:1997nk}, while the
matrix $P$ can be found in \cite{Huiszoon:2001wv}. In the following
we compute for simplicity  the crosscap entropy and the LREE for
diagonal $SU(2)_k$ with $k$ even.   The explicit $P$ matrix for this
case is given in the appendix, although the explicit crosscap states
are given in \cite{Horava:1990ee,Brunner:2001fs}.

We recall that the crosscap entropy is  $S_\mu^C = \ln \Gamma_{\mu
0} = \ln (P_{\mu 0}/\sqrt{S_{00}})$. For the involution ${\cal I}: h
\rightarrow  {h}^{-1}$, the crosscap state is a PSS state (where
$\mu = 0$) and it corresponds to the fixed points of the center of
$SU(2)$:
\be
 S^C_0 = \ln \Big[ \Big(\frac{2}{k+2} \Big)^{\frac{1}{4}} \text{tan}^{\frac{1}{2}}\frac{\pi}{2(k+2)}   \Big] \,.
 \ee
For the other involution  ${\cal I}: {h} \rightarrow - {h}^{-1}$ the
simple current is in the representation $k/2$ with the center
isomorphic to ${\mathbb Z}_2$. The corresponding crosscap
coefficient is given by \eqref{eq:simcurcoe}, then
 \be
 S^C_{k/2} = \ln \Big[ \Big(\frac{2}{k+2} \Big)^{\frac{1}{4}} \text{cot}^{\frac{1}{2}}\frac{\pi}{2(k+2)}   \Big] \,.
 \ee
It is interesting to note that for $S^C_0$, the crosscap ground
state degeneracy  $g_0 <1$  due to the fact that  $ {\hat
\D}/\sqrt{\D} > \hat{d}_0=1$, while for  $ S^C_{k/2}$  the ground
state degeneracy $g_{k/2} >1$  since  ${\hat d}_{k/2} >
{\hat\D}/\sqrt{\D}$ where the expressions for these quantities are
given in \eqref{eq:wzwqd}.


  The central charge in these models is $c = k/(k+2)$, the LREE are
\be
  S_{|C_{0}\rangle} = \frac{\pi \ell c}{6 \varepsilon} - \sum_j \frac{4}{k+2} \, \text{sin}^2 \,\frac{\pi(2j+1)}{2(k+2)}\, \ln \Bigg[  \sqrt{\frac{2}{k+2}} \text{tan}\, \frac{\pi(2j+1)}{2(k+2)}  \Bigg]\,.
 \ee
  For the standard parity and for the non-standard one has
  \be
  S_{|C_{k/2}\rangle} = \frac{\pi \ell c}{6 \varepsilon}  - \sum_j \frac{4}{k+2}\,  (-1)^{2j} \text{cos}^2\, \frac{\pi(2j+1)}{2(k+2)} \,\text{ln} \Bigg[
   \sqrt{\frac{2}{k+2}} (-1)^{2j}\text{cot}\, \frac{\pi(2j+1)}{2(k+2)}
  \Bigg].
  \ee

\section{Final Remarks}
\label{sec:final}

LREE has been exhaustively studied for $2d$ theories defined on
oriented surfaces. In this context it is known that LREE of the edge
theory is deeply connected to entanglement entropy of the 2+1 bulk
theory. However, very few is known on LREE on non-oriented surfaces.
The extension of the edge/bulk correspondence mentioned in the
oriented case, to the non-oriented case is also unknown. Therefore
we find interesting to study the crosscap entropy (which is
explained in the article) and  LREE for crosscap states and their
interrelations with other entropies, as the topological entropy. We
hope that our results bring some light into these directions.  In
what follows we describe some of our results.

A relation between the topological entropy and the boundary entropy
does not necessarily exist. However, in this work, we have found
that, for some unitary and non-unitary minimal models in the A-A
series, the change of topological order from the identity
representation to the one associated with the boundary is manifested
as a boundary entropy.

For the crosscap entropy  we  can not have such interpretation,  as
the crosscap entropy includes also the matrix $P$. However, we
defiend  $\hat{S}$  as an auxiliary tool to express the crosscap
entropy in a similar fashion as the boundary entropy.  Thus the
crosscap entropy can be written as the difference between $\hat S$
and ${1\over 2} S^{top}$. We could interpret $\hat S$ as a
topological entropy due to the topological order when the system is
defined on a non-oriented surface.


We have also computed the left-right entanglement entropy for
crosscap states that preserve a diagonal subalgebra. Similarly, as
in the boundary case, the leading term in the LREE
(\ref{eq:lreetop}) depends on the underlying CFT and it diverges as
the UV cut-off $\varepsilon$ goes to zero. The subleading terms in
the LREE are the crosscap entropy and the sum over the non-trivial
representations. This was shown explicitly for the case of the
$SU(2)_k$ WZW model.


We want to address some concerns about the explicit
expressions of the crosscap coefficients in some of the minimal
models. In \cite{Maloney:2016gsg,Tsiares:16} the authors  considered
the correspondence between  quantum gravity in  $\text{AdS}_3$ and
$1+1$ non-oriented CFT. In particular, they study pure quantum
gravity (which includes only smooth saddles). Then, they restrict
their analysis to minimal models in the CFT side. They conclude that
in the dual minimal models, the crosscap coefficients $\Gamma_{i0}
=0$  must vanish for operators with conformal dimension $h_i<c/12$
to have dual smooth saddles. They were concerned about the existence
of models with such  condition. Here we have computed the crosscap
coefficients for the Tetracritical model (see \eqref{eq:gammaTetra})
and we found that there is a primary field with $h= 1/40 = 0.025$
and the corresponding crosscap coefficient is $\Gamma_{02} =0$. We
expect that this result could contribute to the understanding of
holography on non-orientable surfaces.

Finally it would be very interesting to develop the formalism of the
topological entropy and its relation to the boundary entropy for the
oriented and unoriented cases for characters given in terms of the
Nahm sum (\ref{nahmSum}) with the precise structure of a mock
modular forms (see for instance, \cite{Dabholkar:2020fde}). We leave
this subject for a future work.

\appendix
\section{Appendix} \label{ap:coefficients}

The matrix $S$ for minimal models in the A-A series is given in
\cite{Itzykson:1989sy,DiFrancesco:1997nk}. The matrix $P$ can be
constructed by means of its definition in terms of the matrix $S$
and $T$, although an simple expression for it has been given in
\cite{Bianchi:1991rd}.  For the  Ising model the crosscap
coefficients are  \cite{Lewellen:1991tb} \be \label{eq:IsingCoef}
\Gamma_{0 0} = \sqrt{\frac{2 +
        \sqrt{2}}{2}}\,,  \hspace{0.5cm} \Gamma_{0 \varepsilon} = \sqrt{\frac{2-\sqrt{2}}{2}} \,,  \hspace{0.5cm}  \Gamma_{0 \sigma}= 0 \,.
\ee
 In this case there is only a crosscap
corresponding to the parity symmetry ${\P}_0$ \be |C_{\P_0}\rangle =
\Gamma_{0 0} \vert C_0 \rangle\!\rangle + \Gamma_{0 \varepsilon}
\vert C_\varepsilon \rangle\!\rangle \,, \ee

From \eqref{eq:lreetop} the LREE is \be \label{eq:isingcol}
S_{|C_{\P_0}\rangle} =\frac{\pi \ell }{12 \varepsilon} -
\frac{2+\sqrt{2}}{4} \text{ln} \Big(\frac{2 +
        \sqrt{2}}{2}\Big) -\frac{2-\sqrt{2}}{2}
\text{ln} \Big(\frac{2 -
    \sqrt{2}}{2}\Big) \ee

For the Tricritical model the primary fields  the crosscap coefficients are
\begin{align}
\label{eq:crosscoefTri}
\Gamma_{00} &= \frac{  \sqrt{2 + \sqrt{2}}\, s_1}{\sqrt{s_2}}\,, & \Gamma_{0 \epsilon}& = \frac{ \sqrt{2-\sqrt{2}} \,s_2} {\sqrt{s_1}} \,,\nonumber \\
\Gamma_{0 \epsilon^{\prime}}& = \frac{ \sqrt{2+\sqrt{2}} \,s_2} {\sqrt{s_1}}\,, & \Gamma_{0 \epsilon^{\prime \prime} }& = \frac{ \sqrt{2-\sqrt{2}} \,s_1} {\sqrt{s_2}}\,,\nonumber\\
\Gamma_{0 \sigma} & =0 \,, &  \Gamma_{0 \sigma^{\prime}} &=0  \,.
\end{align}

These results agree with the quotients between them  found using the
sewing constraints in \cite{Tsiares:16}.

The crosscap state is \be |C_{\P_0}\rangle = \Gamma_{00} \vert C_0
\rangle\!\rangle + \Gamma_{0 \epsilon} \vert C_\varepsilon
\rangle\!\rangle + \Gamma_{0 \epsilon^{\prime}}\vert
C_{\varepsilon^{\prime}} \rangle\!\rangle  + \Gamma_{0 \epsilon
^{\prime \prime}}\vert C_{\varepsilon^{\prime \prime}}
\rangle\!\rangle\,, \ee
The corresponding LREE is given by \be
S_{|C_{\P_0}\rangle} = \frac{7\pi \ell }{60 \,\varepsilon} - 4
\Big[s_2^2\, \mbox{ln}\Big(\frac{s_2^2}{s_1}\Big) -
s_1^2\,\mbox{ln}\Big(\frac{s_1^2}{s_2}\Big)\Big]+(s_1^2+s_2^2)
\Big[(2-\sqrt{2})\mbox{ln}(2-\sqrt{2}) +
(2+\sqrt{2})\mbox{ln}(2+\sqrt{2}) \Big] \,. \ee

For the Tetracritical model we label the different conformal fields
with $i = 0, \ldots 9$ it the order given for the conformal weights
in \ref{subsec:oriented}. The crosscap coefficients are:
\begin{align}
\label{eq:gammaTetra}
\Gamma_{0 0} & = \frac{\sqrt{3}+1}{3^{1/4}}\frac{s_1}{\sqrt{s_2}}\,, & \Gamma_{0 1} & = \frac{\sqrt{3}-1}{3^{1/4}}\frac{s_2}{\sqrt{s_1}}\,, & \Gamma_{0 3} & = \frac{\sqrt{3}+1}{3^{1/4}}\frac{s_2}{\sqrt{s_1}}\,, \nonumber\\
 \Gamma_{0 5}&= \frac{\sqrt{2}}{3^{1/4}}\frac{s_2}{\sqrt{s_1}}\,, & \Gamma_{0 6} & = \frac{\sqrt{3}-1}{3^{1/4}}\frac{s_1}{\sqrt{s_2}}  &  \Gamma_{0 8}&= \frac{\sqrt{2}}{3^{1/4}}\frac{s_1}{\sqrt{s_2}}\,,\nonumber\\
    \end{align}
and  $\Gamma_{0 2}= \Gamma_{0 4} =  \Gamma_{0 7}= \Gamma_{0 9}=0$.

The crosscap state and its corresponding LREE can be obtained also
for the Tetracritical model. This can be performed in a
straightforward way and  will not be carried out here.

The matrix $P$ for the $SU(2)_k$ WZW model for $k$ even is \cite{Brunner:2001fs}, then
\be
P_{jl}= \frac{2}{\sqrt{k+2}} \text{sin} \frac{\pi(2j+1)(2l+1)}{2(k+2)} \,, \quad \quad  j+l \in \mathbb{Z}
\ee

\begin{align}
\label{eq:wzwqd}
{\hat \D} &= \frac{1}{P_{00}}= \frac{\sqrt{k+2}}{2} \text{csc}\Big(\frac{\pi}{2(k+2)}\Big) \,,& {\D} &= \frac{1}{S_{00}}= \sqrt{\frac{k}{k+2}} \text{csc}\Big(\frac{\pi}{k+2}\Big) \nonumber\\
{\hat d_j}& = \frac{P_{0j}}{P_{00}}= \text{csc}\Big(\frac{\pi}{2(k+2)}\Big) \text{sin}\Big( \frac{(2 j +1)\pi}{2(k+2)} \Big) & &
\end{align}

 \vspace{.5cm}
\centerline{\bf Acknowledgments} \vspace{.5cm} We are grateful to
Leopoldo Pando-Zayas for reading the manuscript and for his useful
comments about it. N. Quiroz would like to thank Cinvestav, IPN for
hospitality.  It is a pleasure to thank the anonymous referee for
her (his) careful reading of the original manuscript and for all the
suggestions provided, which helped us to improve our article.


\end{document}